# Active interfacial dynamic transport of fluid in fibrous connective tissues and a hypothesis of interstitial fluid circulatory system

Hongyi Li, [1] Yajun Yin, [2] Chongqing Yang, [1] Min Chen, [1] Fang Wang, [1] Chao Ma, [3] Hua Li, [4] Yiya Kong, [1] Fusui Ji, [1] Jun Hu [5]

[1] Beijing Hospital, National Center of Gerontology, Beijing, China

[2] Department of Engineering Mechanics, Tsinghua University, Beijing, China

[3] Institute of Basic Medical Sciences, Chinese Academy of Medical Sciences, Department of Human Anatomy, Histology and Embryology, Neuroscience Center; School of Basic Medicine, Peking Union Medical College, Beijing, China

[4] Institute of Computing Technology, Chinese Academy of Sciences, Beijing, China

[5] Key Laboratory of Interfacial Physics and Technology, Shanghai Institute of Applied Physics, Chinese Academy of Sciences, Shanghai, China; Shanghai Synchrotron Radiation Facility, Shanghai Advanced Research Institute, Chinese Academy of Sciences, Shanghai, China

**Corresponding author:**

Hongyi Li, Cardiology Department, Research Center for Interstitial Fluid Circulation & Degenerative Diseases and Aging, Beijing Hospital, 100730 Beijing, China. Email: leehongyi@bjhmoh.cn;

**Abstract**

Fluid in interstitial spaces accounts for ~20% of an adult body weight. Does it circulate around the body like vascular circulations besides a diffusive and short-ranged transport? This bold conjecture has been debated for decades. As a conventional physiological concept, interstitial space was the space between cells and a micron-sized space. Fluid in interstitial spaces is thought to be entrapped within interstitial matrix. However, our serial data have further defined an interfacial transport zone on a solid fiber of interstitial matrix. Within this fine space that is probably nanosized, fluid can transport along a fiber under a driving power. Since 2006, our imaging data from volunteers and cadavers have revealed a long-distance extravascular pathway for interstitial fluid flow, comprising four types of anatomic distributions at least. The framework of each extravascular pathway contains the longitudinally assembled and oriented fibers, working as a fibrous guiderail for fluid flow. Interestingly, our data showed that the movement of fluid in a fibrous pathway is in response to a dynamic driving source and named as dynamotaxis. By analysis of some representative studies and our experimental results, a hypothesis of interstitial fluid circulatory system is proposed.

**Introduction**

This paper will review some studies on interstitial fluid (IF) flow and summarize our experimental discoveries in active interfacial dynamic transport of IF. This active IF transport is of distinguished characteristics below: (a) It is long-ranged instead of short-ranged. (b) It is systemic instead of regional. (c) It is ordered instead of random. (d) It is structuralized, multi-layered, multi-leveled, multi-scaled, fiber-oriented and interfacial-constrained. (e) It is power-controlled, i.e. heart-beating-driven or periodic-power-source-driven. (f) It is a faster transport process rather than a slower diffusion process, i.e. the measured peak flowing speed is approximately 2 cm/sec. (g) It is systemically cycled, i.e. the exchanges with vascular circulations and the dynamic cycled patterns are universally exist. Finally, a hypothesis of interstitial fluid circulatory system is proposed.

**1. The traditional viewpoints about IF flow**

The total body fluids in adult can be divided into three compartments: (a) the fixed intracellular fluid compartment that accounts for about 32% of body weight, (b) the systemically circulatory fluid compartment (blood plasma, lymphatic fluid and cerebrospinal fluid) for about 8% of body weight, and (c) the local-transport interstitial fluid (IF) compartment for about 20% of body weight. It is noted that the word "local-transport" or "diffusive transport" is used when IF is mentioned in literatures.

In physiology, IF is thought to be entrapped within the interstitial/extracellular matrix [1]. Generally, the composition of interstitial matrix comprises two phases: the fluid/gel phase, consisting of the combination of proteoglycan filaments and IF entrapped within them, so called tissue gel, and the solid phase, consisting of a three-dimensional cross-linked network of fibers embedded within tissue gel. The interstitial matrix in biology is usually regarded as porous materials, and the IF transport through the fluid/gel phase is believed to obey the Darcy's law. When the concentration gradient or pressure gradient is present, IF randomly and diffusively transports water and solutes in a short-ranged space among cells, blood capillaries and initial lymphatic vessels to modulate cellular environment in tissues. Although almost all IF is normally entrapped within tissue gel, a part of IF can still be free of the proteoglycan molecules and flow randomly and locally in the short-ranged regions as well. The following phenomenon has been recorded in the textbook of physiology: when a dye is injected into the circulating blood, it often can be seen that IF flows freely through interstitium, usually coursing along the surfaces of collagen fibers or the surfaces of cells [2].

Roughly speaking, a few key words about the characteristics of traditional IF diagram may be summarized: entrapped, localized, short-ranged, disordered, permeated, single-level structured, single-size scaled, concentration-gradient driven or pressure-gradient driven, Darcy's law controlled. However, such localized diagram of IF flow has been debated for several decades.

## 2. Representative studies related to IF transport in the history

Could the IF transport systematically throughout the whole body? As far as we know, there are three representative studies on the IF transport beyond vascular circulations in history.

The earliest record of IF is "the fluid in perivascular spaces" described in 1851 by Rudolph Virchow, which is an extravascular channel for IF flow distributed between the outer and inner lamina of brain vessels or around fenestrated capillaries. This diagram is further studied recently in researches on the clearance pathways of the brain. A perivascular drainage of IF along the basement membrane of smooth muscle cells of intracerebral and cerebral arteries and capillaries has been observed by identifying the deposits of amyloids [3]. A paravascular drainage pathway of IF along arterial and venous walls has been assessed by injection of labeled Aβ in brain [4]. By rapid distribution of tracer protein throughout the brain from the subarachnoid space, fluid circulation through central nervous system via paravascular pathways has been proposed [5]. Basically, the perivascular and paravascular pathways are regarded as a channel around blood vessels. Their topographical connections with the heat, the directions of flow, the histological structures of the perivascular and paravascular spaces, their exact relationship with the arteries, veins and capillaries are elusive.

The second is the interstitial tissue channel. Around the 1970's, by injection of ferrocyanide ions into blood vessels from which they leak into the interstitial tissues, it was found that the precipitate deposits of ferric ions were in the connective tissues near vessel wall or in the skeletal muscle tissues under the electron microscope, light microscope and in the mesentery of rabbit and cat by darkfield transilluminations [6-8]. To understand the transport of the fluid and the ionic tracer through the extravascular interstices in fibrous connective tissues, a water-rich region in extravascular connective tissues and named it as "prelymphatic tissue channel" described by G. Hauck or "interstitial tissue channel" by Casley-Smith. The interstitial tissue channel forms a converging drainage system from the arterial side of capillaries towards their venous sides and the initial lymphatics nearby. In some tissues like brain and retina where is lacking in lymphatics, the interstitial tissue channels can be very long and perform a function of the prelymphatic pathways from the deep portions of brain tissues to the lymphatic vessels in the neck. However, the tissue channels and their surrounds cannot be stained quantitatively and cannot be directly visualized by electron microscope. Until now, the detailed transport process and the space structures of IF flow through the gel-like interstitial matrix has not been fully understood.

The third is the explorations on the Meridian and Collateral Channel network described in traditional medical literatures. Around the 1980's, instead of intravenous injection, an isotopic tracer is hypodermically injected into an acupuncture point in hands or feet of humans [9]. The centripetal transport processes of the isotopic tracer from the hypodermic injection points are clearly visualized under the single-photon emission computerized tomography (SPECT). The isotope trajectories suggested an extravascular pathway originated from an acupuncture point in the extremities. Another similar phenomenon of the long-distance IF flow from an acupuncture point has been studied by the injection of Trypan blue or Alcian blue and attributed to "Bonghan ducts" or "Primo-vascular system", through the vascular conduits of which, IF can flow [10]. However, the microscopic structures of primo-vascular-conduit is unclear yet.

The above experiments were heuristic: (a) To visualize the extravascular IF transport pathways, the imaging tracer shall be injected into a certain region of interstitial tissues outside the vascular vessels. (b) The strategy to identify a tubular structure for fluid flow may be ineligible. A few essential questions need answered: (a) How to clarify the anatomical and histological structures of the long-distance IF transport pathways in humans? (b) What are the mesoscopic and microscopic structures conducting the IF flow? (c) What is the transport mechanism of the long-distance IF flow? (d) What is the role of the beating heart in IF flow?

## 3. Our experimental discoveries about the systemic IF transport

Since 2006, our research group has made a series of progresses exploring the IF transport. Different from the isotopic tracers, we use a number of techniques with higher spatial resolution including contrast-enhanced MRI and have visualized a long-distance extravascular transport pathway from a hypodermic point in the extremity endings of volunteers, usually an acupuncture point [11]. After the hypodermic injection of Gd-DTPA into the acupuncture points in the hand and foot of human subjects, two types of the long-distance extravascular pathways were enhanced by the paramagnetic tracer in the extremities: the pathways with unsmoothed continuous trajectories and the pathways with smooth continuous trajectories. If the injection point on the hand or foot is in the vicinity of a venous vessel but not an acupuncture point, only the smooth pathways could be displayed. Further analysis of the imaging data revealed that the smooth pathways seemed to have enhanced partial walls of the venous vessel other than the intravascular lumen. The unsmoothed pathways from acupuncture points were in the subcutaneous tissues and have a characteristic of "puncture resistance from acupuncture needle". Through lymphangiography by the hypodermic injection of iodized oil into an acupuncture point, we confirmed that neither the smooth nor the unsmoothed pathways were not in accordance with the visualized lymphatic vessels [12].

## 3.1 The anatomical and histological structures of an extravascular pathway from an acupuncture point in humans

Another valuable progress from our research group is the discovery of the anatomical and histological structures of the long-distance extravascular pathways from an acupuncture point, especially in physiological conditions. One subject was recruited who had severe gangrene foot due to arteriosclerosis obliterans and would receive the amputation of his right lower leg. BEFORE the amputation, he took the hypodermic injection of the fluorescent tracer into an acupuncture point in ankle. Around 90min after the amputation, tissue samples from the lower leg were histological analyzed. It was clearly visualized by the fluorescein that four types of the extravascular pathways from the ankle to the amputated end of leg were consistently formed by fibrous connective tissues [13]: (a) a cutaneous pathway in the dermis and hypodermis tissues, (b) a perivenous pathway along the venous adventitia (including the adventitia), (c) a periarterial pathway along the arterial adventitia (including the adventitia), (d) a fibrous-endoneurium-perineurium-epineurium pathway in the nerves. These results clearly demonstrate that the IF transport pathways are of multiformity and have at least four types of anatomical distributions. Conventional concept of tissue channel is insufficient to sum up such multiformity. We conjecture that the smooth pathways observed in volunteers by MRI were probably the perivascular pathways and the unsmoothed pathways were the cutaneous pathways.

In addition to physiological conditions, IF transport pathways in non-physiological conditions were also investigated. Three amputated lower legs received the injection of the fluorescent tracer into the same acupuncture point AFTER their amputation were investigated. In these cases, periodic "mechanical compressions" on the amputated end of the legs were covered by a sphygmomanometer cuff with a systolic pressure of 50-60 mmHg and a compression-relaxation frequency of 18-20 times/min. After around 90 min of manipulation, the same four types of the fluorescently stained fibrous pathways were observed [13].

The IF transport pathways discovered above in both physiological and non-physiological conditions may lead to the following judgments: (a) The long-distance IF transports include at least four types of anatomical structures in living human body or even in the cadavers. (b) The histological structures of each IF pathway are consistently formed by fibrous connective tissues which suggest that IF flow universally exists in fibrous matrices throughout the whole body. (c) The driving powers of long-distance IF transport were not fully understood yet but might be periodically power-sources-driven or multi-driving-centers-driven. The to-and-fro movements of the dynamic source is of vital importance in IF transport, like the pulsation of the heart.

We have disclosed some optical imaging features of the long-distance IF transport as well. Using two-photon confocal laser microscopy (TPCLSM), all the extravascular pathways were found to contain thousands of the fluorescently stained and micron-sized fibers distributed longitudinally along the long axis of the transport pathways [13]. The following findings may be concluded: (a) The stained fibers are the results of the fluorescent IF flow through the extravascular pathways and there must be a fluid film on a fiber. (b) The space on the fiber may be surrounded by the gel substances and there must be a constrained space for IF film flow along a fiber. (c) The longitudinal fibers (or fibril-bundles) play a role as the guiderail for IF flow in the oriented interstitial connective tissues. (d) Because the distributions of the fibers are long-ranged and oriented, the IF transport is long-ranged and ordered accordingly. In short, as a fibrous guiderail (fibrorail), the fibers are of essential importance in IF transport through a highly structuralized interstitial matrix.

To further confirm the universal existence of the above discoveries in the whole body, the human cadavers were used to testify the extravascular pathways for IF transport by the simulated heart pulsation using a mechanical chest compressor [14]. Upon the hypodermic injection of the fluorescent tracer into an acupuncture point in the first knuckle of thumb, a fluorescently stained extravascular pathway from the right thumb to right atrium near chest wall were visualized after 2.5-hours of repeated chest compressions. The cutaneous pathways from the thumb were found in dermic, hypodermic and fascial tissues of the hand and the lower forearm, but not found in the skin above the level of fossa cubitalis. The perivascular pathways from the thumb were observed along the veins of the arm, axillary sheath, superior vena cava and in the superficial tissues on the right atrium. Histological and micro-CT data showed that these long-distance pathways were neither blood nor lymphatic vessels but the fibrous connective tissues, in which the micron-sized fibers were longitudinally assembled from the thumb to the superficial tissues on the right atrium and appendage. These anatomical data verified that the structural framework of fibrous connective tissues is composed of the multi-layered, longitudinally assembled and cross-linked micron-sized fibers, which provide a fibrorail network to guide IF flow. By TPCLSM, it was clearly observed that these connective tissues comprise the abundant fluorescently stained fibers oriented along the long axis of the transport pathway.

**3.2 An interfacial transport zone within interstitial matrices and an interfacial dynamic transport pathway at a multi-scale**

In interstitial connective tissue, neither the fiber nor the gel themselves can flow. Based on the above experimental evidences at macroscopic and mesoscopic scales, we can construct the IF transport space as follows: The fluorescently stained micron-sized fibers in fibrous matrix are an *in situ* imaging evidence of a fluid film paved on the fiber by the optical microscope. There are no other visible features in this transport region except for the visualized fibers and gel under the submicron resolution. From the viewpoint of interface science, an interfacial space would be formed physically between two phases of matters. In interstitial matrix, the only possible transport space for IF flow must be an *interfacial transport zone* (**ITZ**) between the fiber and the gel/liquid substance [15]. The ITZ is not an interface of a few molecular layers but contains a thin, probably a nanometer-thickness space paved on a solid surface [14]. The space shape of an ITZ mainly depends on the solid surfaces of a fiber, a cell, a bundle of fibers, or a group of cells, and may be variform in a biological system. In a transverse section view, an ITZ has two types of the interfaces: an interface on a solid surface of fibers or cells, and an interface in contact with the gel phase of matrix. The pore sizes or compartment sizes of the ITZs might be variable in physiological or pathophysiological conditions. Indeed, the detailed microscopic structures of the ITZ need further studies.

In macroscopic scale like the anatomical structures, the microscopic ITZs can be topologically connected along the intrinsic fibers and comprise a macroscopic transport pathway of long-ranged characteristics throughout the whole body. When IF comes into the topologically connected ITZs, IF may continuously flow via interstitial connective tissues by a driving power and present two types: IF flow along the embedded, oriented and ordered fibrorails and in the meanwhile diffuse/exchange between the fibers and the surrounding tissue gel along the long-distance extravascular pathways [13]. This IF transport pattern is named "*interfacial dynamic transport* (**IDT**)". According to the anatomical positions, the IDT pathways can be classified into different types, such as a cutaneous IDT pathway, a neural IDT pathway, a facial IDT pathway, an adventitial IDT pathway, an extracellular IDT pathway, a tumorous IDT pathway, a neoplastic IDT pathway, an interosseous IDT pathway, a glandular IDT pathway, etc. [14]. Just like the long journey of vascular circulations since the 17$^{th}$ century, the understandings of an active IDT network need in-depth studies to both the anatomical and microscopic structures and functions of the systemic IF flow and is critical to comprehend fluid circulations in life system.

### 3.3 Imaging of the adventitial IF flow *in vivo*

In rabbits, a continuous fluid flow in both the adventitial pathway and the surrounding fibrous connective tissues as well as their topographical connections with the heart has been observed *in vivo* [16]. Visualized by the fluorescent tracer that was injected into the interstitial spaces around the arterial and venous vessels in the ankle, the fluorescent IF from the ankle was driven along the adventitia and its surrounding fibrous connective tissues of vasculature, into the atrioventricular, anterior and posterior interventricular grooves of the heart, forming pericardial fluid in the rabbits. At the same time, the peripheral fluorescent IF was found in the walls of segmental small intestines and right pulmonary veins as well, which indicated that there was probably more than one center of driving powers in physiological conditions. The measured speed of venous adventitial IF flow was around 0.2-2 cm/sec in the rabbit [16]. By Doppler technique, the measured speed of femoral arterial blood flow in the rabbit was around 5-50 cm/sec and the femoral venous blood flow was around 6-7 cm/sec [17]. Usually, the diffusion speed from a capillary to a cell is around 1 mm/sec [18]. As far as we know, the visualized IF flow is faster than diffusion and seems to fill in the gap of velocity spectrum for body fluid from a lowest diffusion to a fastest arterial blood flow.

A bulk-like emerging flow in the adventitial IDT pathway has been dynamically recorded by fluorescence microscope in *in vivo* rabbits in our lab. Firstly, when an incision was made in a segmental venous adventitia, continuous IF flow can be observed in the incision area and exhibit a continuous bulk-like emerging flow of IF in the ITZs along the fibers of the adventitial IDT pathway. Secondly, when the fluorescein in the IDT pathways began to fade away, a few fluorescent plaques formed by dozens to over one hundred micrometers were found to flush away, that probably represented a bigger ITZ on layers of smooth muscle cells between the tunica adventitia and tunica media of vascular vessels. These imaging data are a distinguished phenomenon of fluid flow via multi-sized and multi-form ITZs in living animals, which may comprehend our understandings on the transport functions of each layer of vascular vessels [3].

### 3.4 The adventitial IF flow exchanges with blood circulation via capillaries

Is the long-distance continuous IF flow cycled? According to classical physiology, the IF is generated by the filtration of plasma from arterial capillaries and retaken by venous capillaries. Our data in the rabbit demonstrated that the long-distance adventitial IDT pathways played a role to deliver the peripheral IF into the pericardial cavity to form pericardial fluid and spread/exchange IF into the surrounding tissues along vascular tree [16]. At the same time, the fluorescent tracer from the IDT pathways was found to stain the venous valves inside the inferior vena cava, which indicated that the fluorescent tracer entered the blood circulation eventually. By tracing the whereabouts of the tracers in the adventitial IDT pathways under the fluorescence microscopy, it was easily found that the fluorescent IF in an IDT pathway has been retaken by the capillary net and converged into a venous vessel nearby. The results coincide with the understanding of IF exchange with capillaries that was described in 1896 by Ernest Starling. Thus, the continuous IF flow can exchange constantly with blood circulation via the capillaries among the long-distance IDT system, including the capillaries in coronary circulation, and circulate around the body.

Most interestingly, our findings suggest vasculature's three kinetic laws for fluid flow: (I) Fluid transport via vascular vessels comprises two types: an intravascular blood flow and an adventitial IDT transport for IF. (II) For venous vessels, the direction of adventitial IF transport is the same as that of blood flow. (III) For arterial vessels, the direction of adventitial IF transport is opposite to that of blood flow. In addition, we define that an adventitial IDT pathway includes three parts (Fig. 1B, 1C): (1) A paravascular/perivascular pathway in the surrounding tissues along vasculature adventitia. (2) An IF transport pathway through vasculature adventitia. (3) An interfacial transport between the tunica adventitia and the surfaces of layers of smooth cells of tunica media. For the arteriole, venule and capillaries in the diverse visceral organs and the brain, the IF flow directions of each part of an adventitial IDT pathway need to be identified. According to our data, the to-and fro movements of the beating heart play a special role as a driving center or transferring center in regulating fluid flow in an adventitial IDT pathway of the vasculature. Within these driving centers, like the heart and the respirometric lung, IF flow in each part of an adventitial IDT pathway need to be identified respectively. The lymphatic vessel, structured like blood vessels, comprises endothelial cells, a thin layer of smooth muscle and fibrous adventitia that binds the lymphatic vessels to the surrounding tissues. We predict that there is an adventitial IDT transport in the lymphatic adventitia as well.

### 3.5 A hypothesis of interstitial fluid circulatory system

In lights of the revealed structural framework of fibrous connective tissues throughout the whole body, the diagram of fluid flow through fibrous matrices are renewed (Fig. 1): (a) The interstitial matrix should not be viewed as a random porous media, but as a three-dimensional network of highly ordered, multilayered and cross-linked fibers embedded with a gel substance. (b) At microscopic scale, fluid flow through fibrous matrix should not only be viewed as a diffusive transport process governed by Darcy's law, but also an interfacial transport process among the *interfacial matrix* driven by an active dynamic mechanism. (c) IF flow is of multiscale characteristic. An ITZ on a fibrorail is microscopic but the topologically connected ITZs on the multilayered, highly ordered and cross-linked fibrorails are macroscopic and long ranged. In an extravascular pathway with a diameter of millimeters and a length of centimeters or even over one meter, fluids in the ITZs would be a bulk-like continuous flow under a driving power. (d) At macroscopic scale or a topographical anatomy, fluid flow through fibrous matrices constitutes an extremely complex network of a long-distance IDT pathway from the extremity endings to visceral organs or among organs or tissues or cells throughout the whole body. (e) The transport processes of the IDT pathway network are "active and dynamic". Without the powers, fluid in matrices would diffuse locally. The movement of fluid in an IDT pathway is in response to a dynamic driving source or center, which is named as dynamotaxis. Usually, the dynamotactic direction of fluid flow in an IDT pathway is toward a driving source or center. (f) It is noted that the driving sources and centers are various other than one unique driving center. The detailed kinetic and dynamic mechanisms to modulate fluid in an IDT system are *waiting-to-be-explored* in an alive biological system and especially humans. (g) Beyond cardiovascular circulation, a network of the IDT pathways may provide a different solution to systemically transport fluids for a very long distance with a fast and vast velocity around the body of animals and humans, carrying oxygen, nutrients, metabolic products, heat, biological substances, extracellular signaling molecules, bio-signals, acupuncture signals, ions and the nanoparticles with sizes of dozens nanometers etc. (h) The IDT system might not only be a transport system but also *an interfacial dynamic communicating system* (**IDC**), communicating and regulating signal exchanges among each part of IDT system by means of need-to-be-understood mechanisms. In our opinion, the IDC system might be modulated systemically also via the continuous daily movements of regular heart beatings, irregular respiration and skin evaporation, etc. (i) Fascinatingly, the reasons of why the heart drives IF into the external tissues of the heart and the whereabouts of IF being transferred by the heart remain unclear. (j) The exact physiological and pathophysiological functions of an active IDT/IDC system need to be comprehensively identified. (k) The electrical potential differences caused by ion flow of interfacial fluid transport via ITZs might generate action potential beyond the membrane of a cell, the electrochemical interactions and physiological roles of which are extremely intriguing.

And above all, the understandings on interstitial fluid are renewed as well (Fig. 2): (1) The interstices of interstitium or interstitial space may refer to a sterical space between the cells. In this bulky space, the majority components of interstitial space are an extracellular matrix that is mainly composed of types of collagenous and elastic fibers, and glycosaminoglycans which form a gel-like substance. In other words, the matrix in interstitial space contains a structural three-dimensional scaffolding embedded in the gel-like interstitium. Firstly, we have further defined a fine space within this structuralized interstitium as an *interfacial transport zone* paved on a solid surface, like a fibrorail or a layer of cells. Fluid in an ITZ can transport along the solid surfaces under an active dynamic driving power, the transport pattern of which is named as an *interfacial dynamic transport*. Secondly, we have revealed the intrinsic structural framework of interstitium is NOT irregular but ordered. By contrast to the bulky and irregular interstitial space between cells and vasculature, the fine ITZ provides a constrained, ordered and oriented space for fluid flow. Molecules and solutes in interstitial spaces would execute not only "random walks" by diffusion but also "oriented transport" along ordered fibrorails. (2) Based on the fundamental definitions of ITZ and IDT/IDC, systemic fluid flow in fibrous matrices may be reconstructed. (3) The conventional definition of interstitial fluid may refer to fluid in the interstitial compartment that accounts for ~ 20% of body weight as well as fluid in the interstitium between the cells and vasculature. Parts of interstitial fluid, which is fluid in an ITZ of interstitial matrices, may be redefined as interfacial fluid. (4) Like the above analysis, interfacial fluid can transport or flow systemically around the whole body at multiple scales, levels and layers, such as an adventitial IDT network, a neural IDT network, a cutaneous IDT network, an facial IDT network, etc. (5) The movements of IF is from and back to blood circulation. Parts of IF enter the ITZs and are transported by an IDT pathway. Fluid in an IDT pathway can quickly and easily interchange with IF and eventually exchange with vascular circulations by capillary beds or other unidentified ways, i.e. the adventitia of an arterioluminal vessel or thebesian veins. (6) Here, we make a hypothesis: (a) Besides diffusive transport in short range, interstitial and interfacial fluid circulate through all parts of the body, which is named as *an interstitial/interfacial circulatory system*, IFCS and abbreviated as *interstitial fluid circulation*. (b) The interstitial and interfacial fluid in extracellular matrices, the blood and lymph in vascular circulations comprise three principal compartments of systemically circulatory system in animals and humans, which together constitute a multiplex circulatory system. (7) Like blood and lymph flow under the actions of the heart, fluid in an adventitial IDT system, which is part of IFCS, is driven by the to and fro movements of the heart as well. (8) Unlike vascular circulations, the topographical structures of fluid flow in the other IDT systems remains unclear, such as the neural IDT system and its driving center, the cutaneous IDT system and its driving center, etc. (9) The comprehensive understandings IFCS and its exact relationships with vascular systems, nervous system and all the other anatomical or functional biological systems need to be further clarified.

Future studies will need an interdisciplinary cooperation to reveal the regulating mechanisms, the relationships with diseases and the detailed anatomical and microscopic structures of IFCS, possibly by a method of dynamic topographical anatomy and a guidance of traditional medicine. The medical applications might benefit from any progress in the understandings on the IFCS, such as medical imaging technique, a novel pathway for drug delivery, therapeutic instruments, etc. Our pilot studies may provoke a new field of fascinating and fruitful researches in life science.

**Statements:**

The authors declare no competing financial interests and have no conflicts to disclose.

**Data Availability Statement:**

The data that support the findings of this study are derived from the public domain resources. Some detailed data that support the findings of this study are also available from the corresponding author.

**Acknowledgements**

This work was financially supported by Beijing Hospital Clinical Research 121 Project (121-2016002) and National Natural Science Foundation of China (11672150).

**Author Contributions**

HyL conceived the original ideas, conceptions and designed the analysis. HyL, QcY, CM, YyK performed the gross human anatomic and animal experiments and the histological analysis. HyL, MC, FW, FsJ, HL analysed the MRI data. HyL, YjY, JH analysed the interfacial transport pattern. HyL wrote the manuscript. All authors contributed to scientific discussions of the manuscript.

**References**

1. Levick J R. Flow through interstitium and other fibrous matrices. Quarterly journal of experimental physiology 1987;72:409-437.
2. Guyton A C. The microcirculation and the lymphatic system: capillary fluid exchange, interstitial fluid, and lymph flow. Textbook of medical physiology 2011:177-190.
3. Bakker EN, Bacskai BJ, Arbel-Ornath M, et al. Lymphatic Clearance of the Brain: Perivascular, Paravascular and Significance for Neurodegenerative Diseases. Cell Mol Neurobiol 2016;36:181-194.
4. Iliff J J, Wang M, Liao Y, et al. A paravascular pathway facilitates CSF flow through the brain parenchyma and the clearance of interstitial solutes, including amyloid β. Science translational medicine 2012:4:147ra111-147ra111.
5. Rennels M L, Gregory T F, Blaumanis O R, et al. Evidence for a 'paravascular' fluid circulation in the mammalian central nervous system, provided by the rapid distribution of tracer protein throughout the brain from the subarachnoid space[J]. Brain research, 1985, 326(1): 47-63.
6. Casley-Smith J R, O'Donoghue P J, Crocker K W J. The quantitative relationships between fenestrae in jejunal capillaries and connective tissue channels: proof of "tunnel-capillaries". Microvascular research 1975; 9:78-100.
7. Casley-Smith J R. Calculations relating to the passage of fluid and protein out of arterial-limb fenestrae, through basement membranes and connective tissue channels, and into venous-limb fenestrae and lymphatics. Microvascular research 1976;12: 13-34.

8. Hauck G, Schröer H. The incomplete darkfield transillumination: a complementary vitalmicroscopic technique. Microvasc Res 1973;6:130-131.

9. Tiberiu R. Do meridians of acupuncture exist? A radioactive tracer study of the bladder meridian. American Journal of acupuncture 1981;9:251-256.

10. Yoo J S, Ayati M H, Kim H B, et al. Characterization of the primo-vascular system in the abdominal cavity of lung cancer mouse model and its differences from the lymphatic system. PloS one 2010;5:e9940.

11. Li H, Yang J, Chen M, et al. Visualized regional hypodermic migration channels of interstitial fluid in human beings: are these ancient meridians? The Journal of Alternative and Complementary Medicine 2008;14:621-628.

12. Li H, Tong J, Cao W, et al. Longitudinal non-vascular transport pathways originating from acupuncture points in extremities visualised in human body. Chinese science bulletin 2014;59:5090-5095.

13. Li H, Yang C, Lu K, et al. A long-distance fluid transport pathway within fibrous connective tissues in patients with ankle edema. Clinical hemorheology and microcirculation 2016;63:411-421.

14. Li H, Yang C, Yin Y, et al. An extravascular fluid transport system based on structural framework of fibrous connective tissues in human body. Cell proliferation 2019:e12667.

15. Li H, Han D, Li H, et al. A biotic interfacial fluid transport phenomenon in the meshwork of fibrous connective tissues over the whole body. Progress in Physiological Sciences 2017;48:81-87.

16. Li H, Chen M, Yang J, et al. Fluid flow along venous adventitia in rabbits: is it a potential drainage system complementary to vascular circulations? PloS one 2012:7:e41395.

17. Qian M, Niu L, Wang Y, et al. Measurement of flow velocity fields in small vessel-mimic phantoms and vessels of small animals using micro ultrasonic particle image velocimetry (micro-EPIV). Physics in Medicine & Biology 2010;55:6069.

18. Herring N, Paterson D J. Overview of the cardiovascular system. Levick's Introduction to Cardiovascular Physiology. CRC Press, 2018:1-2.

Figures and Legends

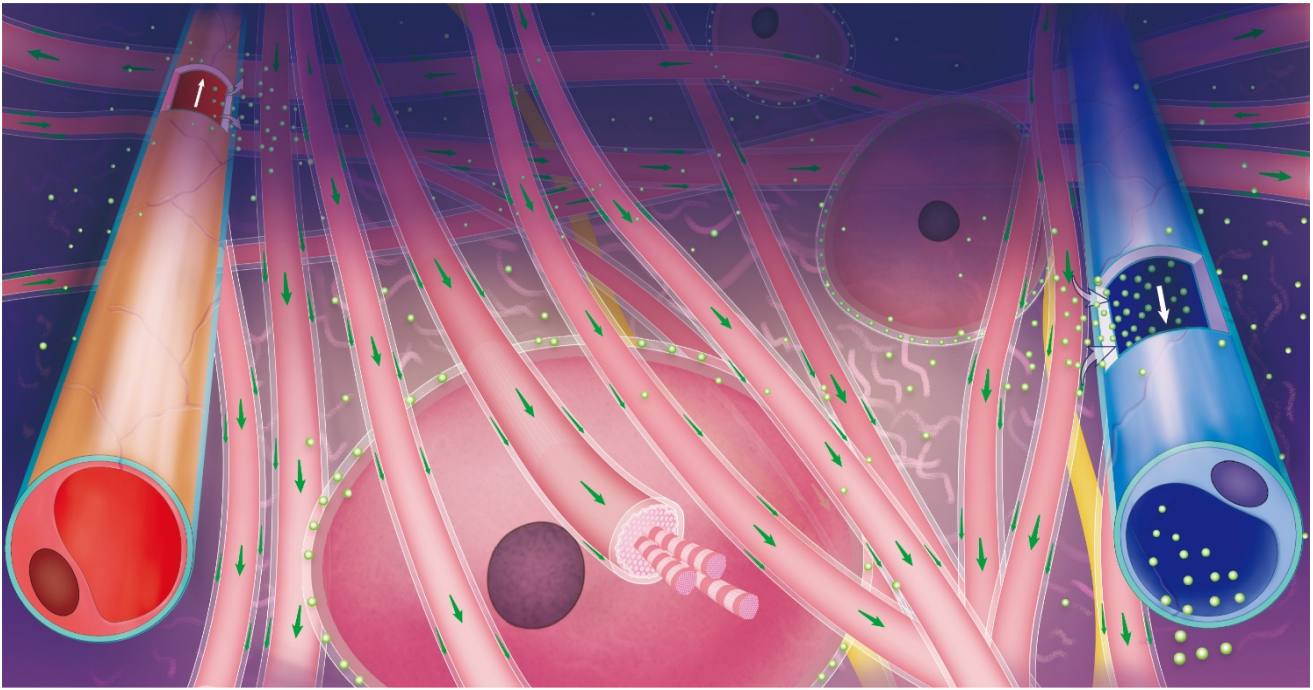

**Figure 1. Illustration of interstitial fluid diffusion and interfacial fluid transport along fibrorails in interstitium.** Derived from an arterial side of the capillary, interstitial fluid (green dots) diffuses into the gel-like substance among fibers and is reabsorbed by the venous side of the capillary, which is a conventional concept of interstitial fluid exchange between the cells and the microcirculation. In the meanwhile, fluid enters into an interfacial transport zone (ITZ) on a fiber and forms interfacial fluid. Interfacial fluid (green arrows) can be transported along the fibrorails of extracellular matrices under an active dynamic driving power. By contrast to the bulky and irregular interstitial space between cells and vasculature, the fine ITZ provides a constrained, ordered and oriented space for fluid flow.

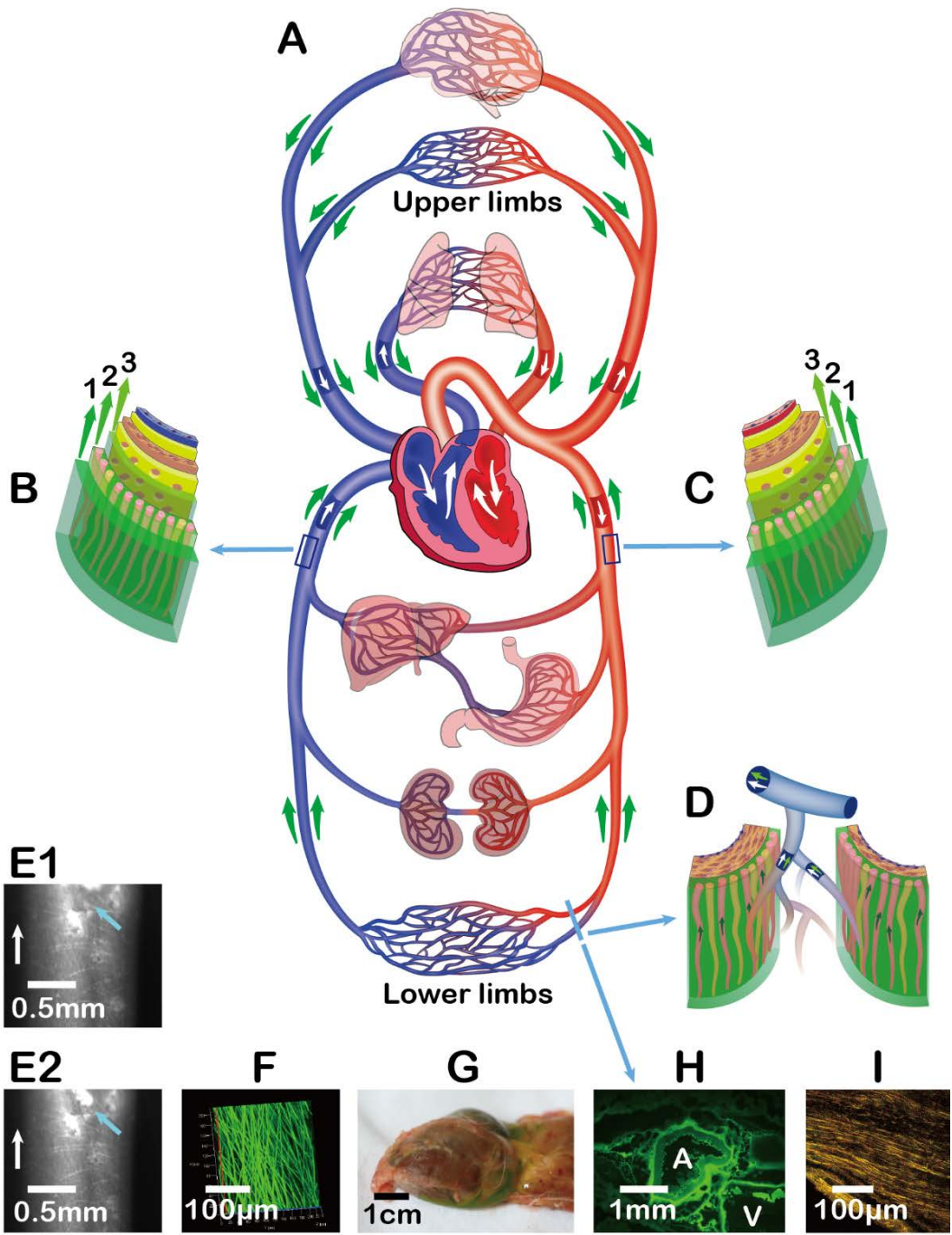

**Figure 2. A diagram of an adventitial IDT system that illustrates part of the hypothesized interstitial fluid circulatory system. A** shows: (I) Fluid transport via vascular vessels comprises two types: an intravascular blood flow (white arrows) and an adventitial IDT transport for IF (green arrows). (II) For venous vessels, the direction of adventitial IF transport is the same as that of blood flow. (III) For arterial vessels, the direction of adventitial IF transport is opposite to that of blood flow. **B** (a

segment of venous wall) and **C** (a segment of arterial wall) show that there are three types of an adventitial IDT pathway for fluid flow: **1**, a paravascular/perivascular fluid flow; **2**, an adventitial fluid flow; **3**, an interfacial fluid flow between the tunica adventitia and tunica media. **D** shows that the fluid in an adventitial IDT pathway is taken by capillaries nearby and converges into vascular circulation eventually. The reabsorption of IF by capillaries occurs in any capillary bed of all parts of the body, including coronary vasculature. **E1** shows a fluorescent plaque of 0.2mm (pointed by a blue arrow) in an adventitial IDT pathway of a venous vessel under a fluorescence stereomicroscope. **E2** shows the fluorescent plaque (blue arrow) flushes away few seconds after **E1**. The findings of **E1** and **E2** strongly suggest there is a bigger interfacial transport zone between the adventitia and media of the vessel, the $3^{rd}$ type of an adventitial IDT pathway. Under confocal microscope, **F** shows the adventitia of the inferior vena cava was stained by the fluorescent fluid from ankle dermis of a rabbit. **G** shows the excessively accumulated pericardial fluid of the heart that was taken an injection of 4-6mL fluorescent fluid into ankle dermis of a rabbit [16]. Under fluorescence microscopy, **H** shows the adventitia and its surrounding tissues of a vein (V) and an artery (A) in the leg were stained by the fluorescent fluid from ankle dermis of a rabbit [16]. Under confocal microscope, **I** shows that the adventitia of an arterial vessel in the amputated leg was stained by the fluorescent fluid from ankle dermis of a human [13].